# Prediction and Realization of a Temperature Control Limit at Low Temperatures in SPRIGT

Haiyang Zhang[a,b,&], Bo Gao[a,b,&,*], Yaonan Song[a,b,&], Changzhao Pan[c,a], Jiangfeng Hu[a,b,e], Dongxu Han[d,a], Ercang Luo[b,a,e], Laurent Pitre[c,a]

[a]TIPC-LNE Joint Laboratory on Cryogenic Metrology Science and Technology, Technical Institute of Physics and Chemistry, Chinese Academy of Sciences, Beijing 100190, China

[b]Key Laboratory of Cryogenics, Technical Institute of Physics and Chemistry, Chinese Academy of Sciences, Beijing 100190, China

[c]Laboratoire national de métrologie et d'essais-Conservatoire national des arts et métiers, F93210 La Plaine-Saint Denis, France

[d]School of Mechanical Engineering, Beijing Institute of Petrochemical Technology, Beijing 102617, China

[e]University of Chinese Academy of Sciences, Beijing 100490, China

[*]Corresponding author, Fax/Tel: +86-10-82554512, E-mail: bgao@mail.ipc.ac.cn

[&]These authors contributed equally to the work and should be regarded as co-first authors.

On May 20th, 2019, World Metrology Day, the Bureau International des Poids et Mesures announced a major revision to the SI in which the definitions of four of the base units were redefined implicitly by fixing the values of four constants of nature [1]. With the help of fundamental constants, one can establish and realize units at the highest level of precision in multiple ways at any place or time and on any scale, without losing accuracy[2, 3]. The practical realizations of the base unit, the kelvin, symbol K, are indicated in the "*Mise en pratique* for the definition of the kelvin in the SI" (*is*-K), in which the absolute and relative primary refractive-index gas thermometry (RIGT) have been included [4, 5]. As an example, single-pressure refractive-index gas thermometry (SPRIGT), the relative primary RIGT, works at a single pressure (*i.e.*, isobar). The unknown thermodynamic temperature of the gas can be determined by comparing the refractive index or resonance frequencies with that at a reference temperature [6, 7]. By using a ratio method, SPRIGT not only reduces the requirement for low uncertainty of pressure measurements and isothermal compressibility but also increases the measured speed of temperatures by about 10 times more rapidly compared with absolute primary thermometry, such as acoustics gas thermometry (AGT) and dielectric constant gas thermometry (DCGT) [6]. Given its high-accuracy and high measurement speeds at temperatures below the neon triple point (24.5561 K), SPRIGT is expected to become widely used.

Temperature stability is a crucial aspect of any state-of-the-art primary thermometry worldwide. To realize accurate SPRIGT measurements, high-stability temperature control is a basic requirement, because the temperature stabilities not only are included in the total uncertainty of the measured thermodynamic temperature but also have a direct influence on microwave and pressure measurements, which are the other two key technologies for SPRIGT. Besides, high-stability temperature control is also necessary for facilitating high accuracy international data comparison below 25 K between TIPC-CAS in China, LNE-Cnam in France, *Physikalisch-Technische Bundesanstalt* in Germany, NPL UK, *Istituto Nazionale di Ricerca Metrologica* in Italy and the National Research Council of Canada in the future. In our previous work, the three key technologies were developed independently: high-stability temperature control, high-stability pressure control and high-accuracy microwave frequency measurement. Temperature stability (standard deviation) of 21 μK - 50 μK was achieved for the pressure vessel using only a single temperature sensor (Cernox CX-1050-CU-HT-1.4L, SN X119000) and a multimeter (Keithley 2002) [8][1], and was later improved to 19 μK over the whole working range by improving the thermal response characteristics of the system [9][2]. Pressure relative stability of several parts in $10^7$ was achieved for the range 30 kPa - 90 kPa using a gas compensation servo-loop at room temperature by maintaining the piston position at a constant height [10]. Microwave resonance frequency relative stability of several parts in $10^{11}$ has been achieved for the quasi-spherical copper resonator under vacuum [7]. Once all the three systems were combined (see Fig. 1a), a rhodium-iron resistance thermometer (RIRT) was used to measure the temperature of the resonator, because RIRTs are the practical thermometers on which realizations of the ITS-90 are most accurately maintained, disseminated and compared [11]. This sensor, (Tinsley, SN 226242), was calibrated by the National Physical Laboratory (NPL) (UK), and is denoted by NPL1. Its resistance is measured using an automatic AC resistance bridge (ASL F900). The stability and accuracy of NPL1 are about 10 times better than those of the Cernox sensor at temperatures from 5 K to 26 K, even the later has a relatively high sensitivity[3]. However, it was difficult to maintain the resonator temperature, measured by NPL1, at the required level by controlling a Cernox sensor. The main reason may be that the resistance of the Cernox sensor was measured by Keithley 2002 multimeter with DC circuits, which leads a slow drift of the measured temperatures. To improve matters, we controlled the resonator temperature instead with a second NPL-calibrated RIRT (Tinsley, SN 226245, denoted NPL2), with a similar temperature coefficient of NPL1 as shown in Fig. 2a (reason see the conclusion of Table 1 below), and a manual AC resistance bridge (ASL F18) set to the desired resistance ratio. The out-of-balance signal is amplified and fed back to the heater to bring the temperature of NPL2 close to the set point [11]. In the manual mode of the F18 bridge, the amplified voltage signal of NPL2 rather than the resistance (temperature) is available and can be used as a feedback for the PID control. In turn, another sensor (NPL1 in the present work) is needed to indicate the temperature stability of the resonator. Fig. 1b plots the simplified schematic diagram of the PID control for the pressure vessel in the present work.

According to the handbooks of both bridges[12, 13], the typical resolution $\delta_T$ (in μK) of the temperature sensor should be (see Supplementary Section 1):

$$\delta_T \approx \frac{0.3\left\{\left(1+\frac{2R_2}{R_t}\right)+\left(1+\frac{2R_1}{R_s}\right)\cdot\frac{R_t}{R_s}\right\}}{\sqrt{\tau}I|\alpha_T|} \quad (1)$$

where $R_s$ and $R_1$ are values of the standard resistor and its potential lead resistance, while $R_t$ and $R_2$ are values of the thermometer resistance and its potential lead resistance. In the present work, two 10 Ω standard resistors (Tinsley 5685A, SN1631808, and SN1580409) were used for F18 and F900 bridges respectively. The variables $\tau$ (in seconds), $I$ (in milliamps) and $\alpha_T$ (in K$^{-1}$) denote respectively the integration time of a single measurement, the excitation current of the AC resistance bridge and the temperature coefficient of the thermometer. The resolution limit is limited by the resistance-temperature characteristics (*R-T* relationship) of the sensor, the voltage noise of the bridges, etc. To realize high-resolution measurements (*i.e.*, with a small value of $\delta_T$), low-noise instruments, high temperature coefficient (absolute value, $|\alpha_T|$) sensors, and low-resistance leads should be used. Once the sensors were installed into the system, the resolution limit mainly depends on the voltage noise of the bridges, which is limited by the fundamental Johnson Noise limit of the bridges.

---

[1] The cryostat has not yet coupled with the resonator. Temperature stabilities of the pressure vessel were investigated at (5.0, 5.7, 7.4, 14.3 and 25) K under vacuum.
[2] The cryostat has coupled with the resonator but not yet with the pressure control system. Temperature stabilities of the pressure vessel were studied at (5 and 25) K under vacuum.
[3] Typical calibrated standard uncertainty $u(T)$ of the RIRTs are 0.25 mK @ 5 K and 0.48 mK @ 26 K; while those of the Cernox 1050 type sensor are 2 mK @ 5 K and 5 mK @ 26 K. Typical sensitivity (d$R$/d$T$) of the RIRTs are 0.43 Ω/K @ 5 K and 0.14 Ω/K @ 26 K; while those of the Cernox 1050 type sensor are -316 Ω/K @ 5 K and -12 Ω/K @ 26 K.

**Fig. 1.** Schematic diagram of cryostat and PID control. (a) Simplified schematic diagram of the cryostat for Single Pressure Refractive Index Gas Thermometry; (b) Simplified schematic diagram of the PID control for the pressure vessel.

The temperature resolution $\delta_T$ is the smallest change that the temperature sensor can detect in the quantity being measured. In practice, with PID control, the smallest stability of the measured temperature (*i.e.*, the temperature stability limit $\sigma_{T,L}$), defined by the standard deviation of the measured temperature corresponding to an integration time of $\tau$, amounts to half the resolution of the temperature sensor $\delta_T$ (see Supplementary Section 2):

$$\sigma_{T,L} \equiv \min(\sigma_T) = \frac{\delta_T}{2} \tag{2}$$

where $\delta_T$ of the two temperature sensors is given by Eq. (1) in the present work. Fig. 2a shows the temperature stability limit $\sigma_{T,L}$ for the two RIRTs for $I = 1$ mA excitation current and $\tau = 33.6$ s. It is apparent that the values of $\sigma_{T,L}$ for NPL2 are similar to but slightly lower than those for NPL1 over the whole range from 4.2 K to 26 K.

To achieve a stable temperature environment at temperatures from 5 K to 26 K for further microwave and pressure measurements in the resonator, a multi-layer radiation shield was added, and a thermal-resistance method and a gas-type heat switch were used. The multi-layer radiation shield and thermal-resistance method were used to reduce the thermal noise from surroundings as shown in our previous work [8]. In our system, the heat switch has two working modes, heat switch on mode, with working gas ($^4$He in the present work) filling the heat switch cavity, and heat switch off mode, without $^4$He in the cavity. Heat switch off mode can reduce the thermal noise from the 2$^{nd}$ cold head of the GM pulse tube cryocooler (Sumitomo SHI-RP-082B2). However, the heat transfer between the resonator and the 2$^{nd}$ flange is weak, thus the pressure vessel temperature can not be cooled down to our minimum objective temperature 5 K (only up to ~ 14.3 K [8]). So the heat switch on mode was used all the time in the present work, with which the resonator temperature can be easily cooled down to ~ 4.3 K [9], although it is less good for the temperature control. Moreover, to reduce the perturbation from surroundings still further, a laboratory was constructed with temperature fluctuations less than ± 0.1 K, and an oil bath (Aikom Instruments MR 5100-L) with temperature stability better than 1 mK was used for the standard resistances to provide more stable measurements. Finally, active PID control was implemented on the pressure vessel based on the schematic diagram of Fig. 1b: 1) set the desired ratio for the controlled thermometer NPL2 with F18 bridge; 2) read the voltage of F18 and get the voltage difference $\Delta V$; 3) heat the pressure vessel to balance the F18 bridge; 3) adjust the PID parameters until the resonator temperature NPL1 is stable, measured by F900 bridge in real-time.

The temperature of NPL1 over a two-hour period was measured at temperatures from 5 K to 26 K and pressures of helium-4 gas in the resonator up to 120 kPa. The excitation currents of the two RIRTs are the same, 1.0 mA. The measurements were firstly implemented from high temperatures to low temperatures under vacuum, then repeated under pressures. The experimental temperature stability results and the predicted temperature stability limit of NPL1 with an integration time of 33.6 s by Eq. (2) are compared in Fig. 2b, where the standard uncertainties of the predicted $\sigma_{T,L}$, $u(\sigma_{T,L})$ in μK, were estimated using the above equations based on error propagation formula. The uncertainty component from the calibrated uncertainty is too small (less than 1 nK), thus the effect can be neglected compared with the stability of μK level. The main uncertainty component for the calculated stability limit is from the lead resistance of the sensors, which means low-resistance lead should be used. We can see that most of the resonator temperature stabilities lie within its limit calculated from Eq. (2), while some points lie outside the calculated stability limits with deviations up to several $u(\sigma_{T,L})$ (out of ± $3u$[4]), especially for higher temperatures. Since the system may not be stable enough under vacuum when the set temperature changed to a new high temperature, it is a little hard to be controlled. As for measurements under pressure, the system is cool enough, thus it is easy to be controlled even at high temperatures. However, there are still some big deviations due to non-optimal PID parameters, which can be improved with care. Besides, all the measurements in the present work were carried out with heat switch on mode, in which the temperature is relatively hard to be controlled compared with heat switch off mode over the whole working range. Also, the resolution of RIRTs at high temperatures are lower than at low temperatures, which is not as good for temperature control. Based on the above analyzation, new measurements under vacuum and 60 kPa have been carried out again and plotted in Fig. 2b with a hollow symbol. These temperature stabilities have been improved by a factor of 2 ~ 3 compared with the previous measured ones, and almost all points are within the calculated stability limits. In the future, we can use heat switch off mode for high temperatures and heat switch on mode for low temperatures, which could make the control more robust.

---

[4] We defined $u = u(\sigma_{T,L})$. The temperature control was considered to be not good enough when the difference between the measured stability and the calculated limit was more than $3u$ (*i.e.*, $(\sigma_T - \sigma_{T,L}) > 3u$).

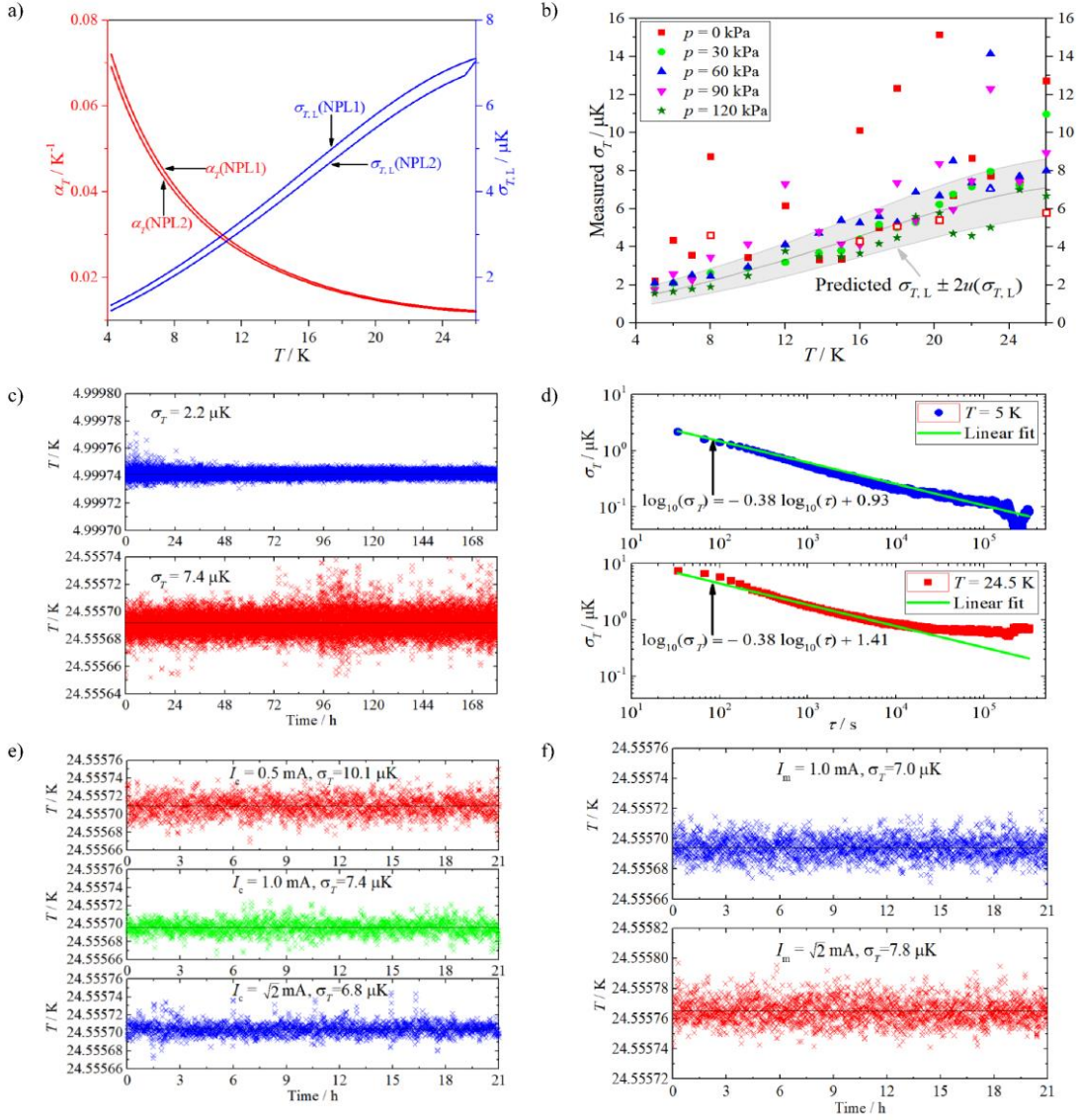

**Fig. 2.** Integration time for the above data $\tau = 33.6$ s and the excitation currents of the two RIRTs $I = 1.0$ mA unless otherwise stated. (a) Temperature coefficient and the temperature stability limit $\sigma_{T,L}$ of NPL1 and NPL2 at temperatures from 4.2 K to 26 K; (b) Comparison between the experimental temperature stabilities of NPL1 and these predicted from Eq. (2) with an uncertainty band of $\pm 2u(\sigma_{T,L})$ (symbol interior: solid, experimental data in the original submission; hollow, new experimental data); (c) 180-hour temperature stability measurement of NPL1 at 5 K and 24.5 K under vacuum; (d) Temperature stability of NPL1 versus integration time at 5 K and 24.5 K under vacuum; (e) Temperature stability of NPL1 response to different NPL2 excitation current under vacuum; (f) Temperature stability of NPL1 responses to different NPL1 excitation current under vacuum.

In a further test, a long-term stability measurement of the resonator temperature for 180 h (7.5

days) was carried out under vacuum at 5 K and 24.5 K with $\tau = 33.6$ s. The measured temperature stabilities of NPL1 are 2.2 µK and 7.4 µK for 5 K and 24.5 K respectively (Fig. 2c). To improve the temperature stability further, one could use a low-noise instrument or high temperature coefficient sensor, or improve the temperature sensor resolution by increasing the integration time or the excitation current, as implied in Eq. (1) (but since the self-heating of the controlled RIRT will increase, to keep the same temperature, the resistance set point should also be increased). Fig. 2d shows a plot of the generalized variance [14] for resonator temperature as a function of integration time for the measurements at 5 K and 24.5 K. Linear fit in logarithmic form was implemented for 5 K and 24.5 K using data with $\tau < 2\times10^5$ s and $1\times10^4$ s, respectively. We can see that NPL1 temperature stability for 5 K is three times better than that for 24.5 K over the whole integration time range, mainly because the relatively higher temperature coefficient at 5 K improves the temperature control stability, as shown in Eqs. (1) and (2). What is more, resonator temperature stabilities of 0.66 µK and 0.10 µK were achieved for 24.5 K and 5 K respectively for an integration time of $\tau = 1.4\times10^5$ s = 1.6 days. This temperature stability lies close to the manufacturer's specified limit for the F900 AC resistance bridge.

    To further check the stability limit, high-resolution temperature sensors were simulated by using high excitation currents on the two AC bridges. The results are summarized in Table 1 and plotted in Figs. 2e and 2f. Fig. 2e shows the temperature stability response of the sensor NPL1, measured with excitation current $I_m = 1.0$ mA, for different excitation currents $I_c$ of the control thermometer NPL2. When $I_c$ was increased from 0.5 mA to $\sqrt{2}$ mA, the temperature resolution of NPL2 was improved, which in turn improved the measured temperature stability of NPL1 from 10.1 µK to 6.8 µK, a value within the limit of NPL1 (6.9±0.7) µK. However, from Table 1 we can see that the 6.8 µK stability lies outside the limit of NPL2 (4.7±0.4) µK with deviation more than $5u$. It means that the temperature resolution of NPL1 for $I_m = 1.0$ mA is not enough to reach the limit of NPL2 for $I_c = \sqrt{2}$ mA if the temperature can be controlled to the limit of NPL2. Fig. 2f shows the temperature stability response of the sensor NPL1 for excitation currents $I_m = 1.0$ mA and $I_m = \sqrt{2}$ mA, while NPL2 was controlled with excitation current $I_c = 1.0$ mA. When $I_m$ was increased from 1.0 mA to $\sqrt{2}$ mA, the measured temperature stability of NPL1 increases from 7.0 µK to 7.8 µK, mainly because of its increasing self-heating. The stability 7.8 µK lies within the limit of NPL2 (6.6±0.6) µK with deviation about $2u$, while it lies outside the limit of NPL1 (4.9±0.5) µK with deviation more than $5u$. It means the stability is also up to the limit of NPL2. The above results further verify that most of the temperature stabilities in Fig. 2b are up to the limit that the two RIRT sensors and the bridges can control and measure. Besides, it also implied that the controlled and measured sensors should have a matched resolution (similar temperature coefficient) to well reflect the temperature stability.

Table 1 Measured temperature stability response to different excitation current at 24.5557 K under vacuum.

| Current $I$ / mA | | Predicted $\sigma_{T,L} \pm u(\sigma_{T,L})$ / µK | | Measured $\sigma_T$ / µK | $Y = (\sigma_T - \sigma_{T,L}) / u(\sigma_{T,L})$ | |
|---|---|---|---|---|---|---|
| NPL2[f] | NPL1[g] | NPL2 | NPL1 | NPL1 | NPL2 | NPL1 |
| 0.5 | | 13.2 ± 1.2 | | 10.1 | -3 < Y < -2 | 4 < Y < 5 |
| 1.0 | 1.0 | 6.6 ± 0.6 | 6.9 ± 0.7 | 7.4 | 1 < Y < 2 | 0 < Y < 1 |
| $\sqrt{2}$[h] | | 4.7 ± 0.4 | | 6.8 | 5 < Y < 6 | -1 < Y < 0 |
| 1.0 | 1.0 | 6.6 ± 0.6 | 6.9 ± 0.7 | 7.0 | 0 < Y < 1 | 0 < Y < 1 |
| | $\sqrt{2}$[h] | | 4.9 ± 0.5 | 7.8 | 2 | 5 < Y < 6 |

[f] Used in temperature control. [g] Used in temperature measurement. [h] Simulated a high-resolution sensor.

In conclusion, methods to estimate and realize the temperature control stability limit in SPRIGT were presented in this work. The resonator temperature stability limit at the micro-kelvin level has been predicted and achieved at temperatures from 5 K to 26 K. The resonator temperature stability can be maintained to better than 8 μK with an integration time of 33.6 s over 180 h, even longer if need be. The experiment results show that the stability is up to the limit that the two RIRT sensors and the bridges can control and measure. In the future, it may be further improved by using low-noise instruments, high temperature coefficient sensors and low-resistance leads. The present work can be used not only in the implementation of high-accuracy SPRIGT, but also in other low-temperature primary thermometry (RIGT, AGT, DCGT, etc.). Besides, it has the potential to estimate the performances of different sensors and instruments. In the future, the present work should provide a solid foundation for international data comparison of thermodynamic temperature at low temperatures, and will promote realizations of the new kelvin and the spread of high-accuracy, low-temperature metrology.

**Conflict of interest**

The authors declare that they have no conflict of interest.

**Acknowledgments**

This work was supported by the National Key R&D Program of China [2016YFE0204200], the National Natural Science Foundation of China [51627809], the International Partnership Program of the Chinese Academy of Sciences [1A1111KYSB20160017], and the EMRP project Real-K [18SIB02]. Author Changzhao Pan was supported by the funding provided by the Marie Skłodowska-Curie Individual Fellowships-2018 [834024]. The authors gratefully acknowledge Richard Rusby from NPL UK for sharing his long experience in calibration and constant advice on the using of RIRT. We are deeply grateful to our colleagues Zhen Zhang and Ying Ma for expert technical assistance. We would also like to thank Dr Mark Plimmer from LNE-Cnam France and Dr Wei Wu from City University of Hong Kong China for their kindly reading and helpful suggestions on the manuscript.

**Author's contributions**

Haiyang Zhang and Bo Gao proposed the method. Haiyang Zhang, Bo Gao, Laurent Pitre, and Ercang Luo conceived and designed the experiments. Yaonan Song, Haiyang Zhang, Changzhao Pan, Jiangfeng Hu, and Dongxu Han performed the experiments. Haiyang Zhang and Yaonan Song analyzed the data and plotted the figures. Haiyang Zhang wrote the paper.

**Appendix A. Supplementary data**

Supplementary data to this article.

# Supplementary Materials for
# Prediction and Realization of a Temperature Control Limit at Low Temperatures in SPRIGT

Haiyang Zhang[a,b,&], Bo Gao[a,b,&,*], Yaonan Song[a,b,&], Changzhao Pan[c,a], Jiangfeng Hu[a,b,e], Dongxu Han[d,a], Ercang Luo[b,a,e], Mark Plimmer[c,a], Laurent Pitre[c,a]

[a]TIPC-LNE Joint Laboratory on Cryogenic Metrology Science and Technology, Technical Institute of Physics and Chemistry, Chinese Academy of Sciences, Beijing 100190, China

[b]Key Laboratory of Cryogenics, Technical Institute of Physics and Chemistry, Chinese Academy of Sciences, Beijing 100190, China

[c]Laboratoire national de métrologie et d'essais-Conservatoire national des arts et métiers, F93210 La Plaine-Saint Denis, France

[d]School of Mechanical Engineering, Beijing Institute of Petrochemical Technology, Beijing 102617, China

[e]University of Chinese Academy of Sciences, Beijing 100490, China

[*]Corresponding author, Fax/Tel: +86-10-82554512, E-mail: bgao@mail.ipc.ac.cn

[&]These authors contributed equally to the work and should be regarded as co-first authors.

## 1. Thermometer resolution

According to the operator's handbooks of F900 [1] and F18 [2], we can know that:

**1)** The typical voltage resolution of the two bridge is $0.3\text{nV}\sqrt{\text{Hz}}$ at 1 ohm matching impedance (page 34 of F18 handbook and page 38 of F900 handbook). *i.e.* the voltage resolution per unit resistance $\delta_v$ (in nV/Ω) is:

$$\delta_v = \frac{0.3}{\sqrt{\tau}}, \text{ (nV/\Omega)} \tag{S1}$$

where variable $\tau$ (in seconds) denotes the integration time of a single measurement. The voltage noise is limited by the fundamental Johnson Noise limit of the bridge.

**2)** The bridge output impedance $R_o$ (in Ω) can be calculated by (page 18 of F18 handbook and page 23 of F900 handbook):

$$R_o = (R_s + 2R_1)n^2 + R_t + 2R_2, \text{ (\Omega)} \tag{S2}$$

where $R_s$ and $R_1$ are values of the standard resistor and its potential lead resistance, while $R_t$ and $R_2$ are values of the thermometer resistance and its potential lead resistance with a unit in ohm; $n$ is the transformer ratio at balance for the bridges with $n=R_t/R_s$. *Note: When measuring low resistances especially for cryogenic applications, the lead resistances are significant and cannot be neglected in the calculation of bridge impedance.*

From Eqs. (S1) and (S2), we can get the voltage resolution $\delta_V$ (in nV) of the measured thermometer as:

$$\delta_V = \delta_v \cdot R_o = \frac{0.3}{\sqrt{\tau}} \times R_o, \text{ (nV)} \tag{S3}$$

For the excitation current $I$ (in milliamps) of the AC resistance bridge, we can simply estimate the resistance resolution $\delta_R$ (in μΩ) by:

$$\delta_R = \frac{\delta_V}{I}, \text{ (\mu\Omega)} \tag{S4}$$

Using the resistance-temperature characteristic of a thermometer, the temperature resolution $\delta_T$ (in μK) can be calculated as follows:

$$\delta_T = \frac{\delta_R}{|dR_t/dT|}, \text{ (\mu K)} \tag{S5}$$

where $dR_t/dT$ (in Ω K$^{-1}$) is the sensitivity of the thermometer.

Combined Eqs. (S2)-(S5), we can determine the temperature resolution $\delta_T$ (in μK) of a thermometer by the following equation:

$$\delta_T \approx \frac{0.3\left\{\left(1+\dfrac{2R_2}{R_t}\right)+\left(1+\dfrac{2R_1}{R_s}\right)\cdot\dfrac{R_t}{R_s}\right\}}{\sqrt{\tau}I|\alpha_T|}, \ (\mu K) \tag{S6}$$

$$\alpha_T = \frac{1}{R_t}\frac{dR_t}{dT}, \ (K^{-1}) \tag{S7}$$

where $\alpha_T$ (in K$^{-1}$) denotes the temperature coefficient of a resistance thermometer. From the above Eq. (S6), we can know that if we want to realize high-resolution measurements (*i.e.*, with a small value of $\delta_T$), low-noise instruments, high temperature coefficient (absolute value, $|\alpha_T|$) sensors, and low-resistance leads should be used.

## 2. Temperature stability limit

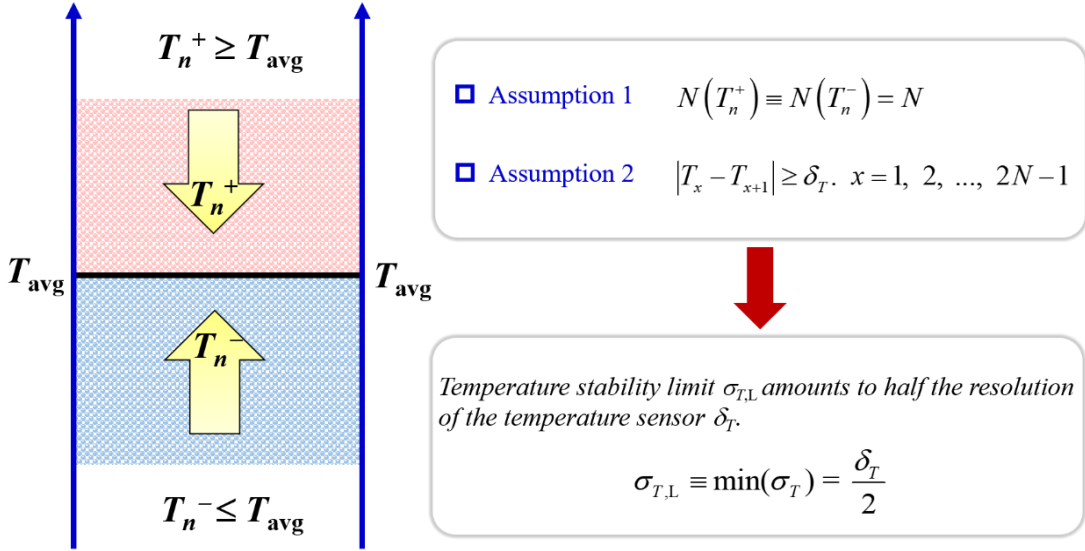

Fig. S1. Simplified schematic diagram for the calculation of temperature stability limit $\sigma_{T,L}$.

In the present work, two assumptions were made as shown in Fig. S1:

**1)** The number of temperatures larger than the average value $T_{avg}$, $N(T_n^+)$, is equal to that of temperatures smaller than $T_{avg}$, $N(T_n^-)$.

$$N\left(T_n^+\right) \equiv N\left(T_n^-\right) = N \tag{S8}$$

$$T_n^+ \geq T_{avg}, \ T_n^- \leq T_{avg} \tag{S9}$$

$$T_{avg} \equiv \frac{\sum_{n=1}^{N} T_n^+ + \sum_{n=1}^{N} T_n^-}{2N} \tag{S10}$$

where $N$ is half of the total measured data point number. By PID control, the measured temperatures are preferred to approach the temperature set point $T_{\text{set point}}$ and usually the measured average value $T_{\text{avg}}$ is very close to or even equal to $T_{\text{set point}}$, *i.e.*,

$$T_{\text{avg}} \approx T_{\text{set point}} \tag{S11}$$

**2)** The difference between two continuous temperature measurements should be not less than the sensor's temperature resolution $\delta_T$.

$$|T_x - T_{x+1}| \geq \delta_T. \quad x = 1, 2, ..., 2N-1 \tag{S12}$$

where $x$ means the data point of the $x$-th measurement.

The stability of the measured temperature, defined by the standard deviation of the measured temperature corresponding to an integration time of $\tau$, can be calculated as follows:

$$\sigma_T = \sqrt{\frac{\sum_{n=1}^{N}\left\{\left(T_n^+ - T_{\text{avg}}\right)^2 + \left(T_n^- - T_{\text{avg}}\right)^2\right\}}{2N-1}} \tag{S13}$$

For $m$ positive numbers $a_i$, we have the following inequalities (*see* page 163 in [3]):

$$\sqrt{\frac{a_1^2 + a_2^2 + ... + a_m^2}{m}} \geq \frac{a_1 + a_2 + ... + a_m}{m}. \quad m = 1, 2, 3,... \tag{S14}$$

where the equality sign is valid only if $a_1 = ... = a_m$.

Combined Eqs. (S9), (S13), and (S14), we can get:

$$\sigma_T \geq \sqrt{\frac{2N}{2N-1}} \cdot \frac{\left(T^+ - T^-\right)}{2} \tag{S15}$$

where the equality sign is valid only if:

$$T_1^+ - T_{\text{avg}} = T_2^+ - T_{\text{avg}} = ... = T_n^+ - T_{\text{avg}} = T_{\text{avg}} - T_1^- = T_{\text{avg}} - T_2^- = ... = T_{\text{avg}} - T_n^- \tag{S16}$$

*i.e.*,

$$\begin{aligned} T_1^+ &= T_2^+ = ... = T_n^+ = T^+ \\ T_1^- &= T_2^- = ... = T_n^- = T^- \\ T^+ + T^- &= 2T_{\text{avg}} \end{aligned} \tag{S17}$$

In this case, Eq. (S17) is also equivalent to say that there are only two measured points $T^+$ and $T^-$. Then from Eq. (S12), we know

$$T^+ - T^- \geq \delta_T \tag{S18}$$

Combined Eqs. (S15) and (S18), we can get the following equation for long-time PID measurements:

$$\sigma_T \geq \frac{\delta_T}{2} \tag{S19}$$

where the equality sign is valid only if Eq. (S17) is satisfied, $T^+ - T^- = \delta_T$, and $N \to +\infty$.

From Eq. (S19), we can know that with PID control the smallest stability of the measured temperature (*i.e.*, the temperature stability limit $\sigma_{T,L}$), defined by the standard deviation of the measured temperature corresponding to an integration time of $\tau$, amounts to half the resolution of the temperature sensor $\delta_T$:

$$\sigma_{T,L} \equiv \min(\sigma_T) = \frac{\delta_T}{2} \qquad (S20)$$

In the present work, temperature sensor resolution $\delta_T$ is described by Eq. (S6). *Note: The above Eq. (S20) is also suitable for other sensors, such as pressure sensor, laser displacement meter.*